\documentclass[letterpaper]{elsarticle}
\usepackage{graphicx,url,natbib,alltt}
\usepackage{subcaption} 

\begin{document}

\title{A New Method for Estimating the Absolute Magnitude Frequency
Distribution of Near Earth Asteroids (NEAs)}
\author[fv]{F. Valdes}
\ead{fvaldes@noao.edu}
\address[fv]{National Optical Astronomy Observatory,
  P.O. Box 26732, Tucson, AZ 85732}

\begin{abstract}

The distribution of solar system absolute magnitudes ($H$) for the near-Earth
asteroids (NEAs) observable near opposition -- i.e. Amors, Apollos, and Atens
($A^3$) -- is derived from the set of \textbf{ALL} currently known NEAs.  The
result is based only on common sense assumptions of uniformly random
distributions and that the orbital phase space and $H$-magnitude
distribution of known NEAs is representative of the total population.  There
is no population or other modeling and no assumption on albedo except in
interpreting the result as a size-frequency distribution (SFD).  The
analysis is based on the 18355 $A^3$ NEAs cataloged by the MPC as of June
2018.  The observations from 9 of the top programs (in terms of number of
distinct NEAs observed) and the smaller but deeper DECam NEO Survey are
used, comprising 74696 measurements of 13466 NEAs observed within 30 deg of
opposition.  The only parameter in the analysis is an estimate of the
detection magnitude limits for each program.

A single power-law slope for the cumulative distribution,
$\log(N\!\!\!<\!\!\!H)=0.50\pm0.03H$, for $H < 27$ is found with no evidence for
additional structure.  A turn-over fainter than 27th magnitude may occur,
but the population of known NEAs is dropping off rapidly because they are
difficult to detect and so possibly is a completeness effect.
Connecting to the nearly complete census of the brightest/biggest NEAs
(diameter $> {\sim}2$Km) provides a normalization that estimates ${\sim}10^8
A^3$ NEAs with $H < {\sim}27$ corresponding to NEAs greater than ${\sim}10$m
in diameter for reasonable typical albedos.

Restricting the analysis to Earth crossing asteroids (10839 known, 7336
selected, 36541 observed) produces the same power-law slope.

\end{abstract}


\maketitle

\section{Introduction \label{introduction}}

The most basic question in the study of near-Earth objects (NEOs) is the
size of the population and how it is distributed by mass.  Answering this
question has important relevance to solar system science.  It also has
ramifications beyond this science domain because of the potential
consequences of impacts with the Earth causing damage, and worse, to the
inhabitants of our planet. This has led to a focus on this basic question by
governments, such as in the U.S. congressional mandate (in the NASA
Authorization Act of 2005 and the detailed study of \citet{Stokes}) to
measure the number of NEOs in the hazardous, if not catastrophic, range --
defined as diameters greater than 140 meters.

The observational version of this question is not the mass or size
distribution but the absolute solar system magnitude, or $H$-magnitude,
frequency distribution (HFD).  Size or mass distributions are derived from this
with considerably more assumptions and interpretation.  This paper estimates
the near-Earth asteroid (NEA) HFD with Amor, Apollo, and Aten ($A^3$) orbits
using a new method based solely on observations that avoids population
models and simulations.

As for all studies, the starting point is the catalog of known near-Earth
asteroids (NEAs).  The catalog used here is that tabulated by the Minor
Planet Center \citep{MPC} as of June 2018.  In order to be in this catalog
an object must have met the MPC criteria needed to determine an orbit and
absolute solar system magnitude as well as the adopted definitions for the
NEA classes.  In addition to orbital parameters, this catalog includes the
individual measurements and attributions to the programs reporting them.

The notable results reported in this paper are 1) an HFD that is a simple
power law and 2) the difference in estimated number of NEAs, in the range of
interest for planetary defense, between this purely observational method and
the studies of \citet{Harris}, \citet{Schunova}, \citet{Granvik}, and
\citet{Tricarico} (henceforth HSGT) using various other methods.  In
particular, the HSGT studies predict a factor of approximately seven fewer
NEAs with diameters larger than the congressional objective than found in
this study.  Clearly this is significant for understanding the feasibility
of mapping these NEAs.

In this paper we describe the method in \S\ref{method} and its application
to the catalog of known NEAs in \S\ref{application}.
Section~\ref{discussion} discusses the results in light of the other
estimates of the HFD and the implication for satisfying the planetary
defense mandate for 140m and larger NEAs.

\section{The Method \label{method}}

The method presented in this paper considers the problem of estimating the
absolute solar system magnitude frequency distribution of near-Earth
asteroids from a purely observational perspective. The dominant factor in
this problem is the volume correction factor; that is, given an observed
sample of NEAs what fraction of the true population was in the observable
volume vs the volume that was unobservable?  In observational methodology
this is known as the Malmquist bias \citep{Malmquist}.

Why is this the dominant factor and not things like detection efficiencies?
Because the concept of near-Earth asteroids, and how they are defined as a
class, focuses on "near by" and the perihelia boundary at 1.3 AU and it is
easy to forget that these asteroids typically have aphelia that put them far
from the Earth during most of their orbit.  This puts them beyond the
current observational limits of the telescopes for considerable periods of
time.

Fortunately, the unobservable population of NEAs can be accounted for
without population modeling and significant assumptions.  This is because
orbital mechanics tells us the probabilities of where NEAs can be
provided we have the orbital parameters.  By definition NEAs have those
parameters in order to be classified as such.  This is the basis of the
methodology presented in this paper.

\begin{figure}[h]
\begin{center}
\includegraphics[width=\textwidth]{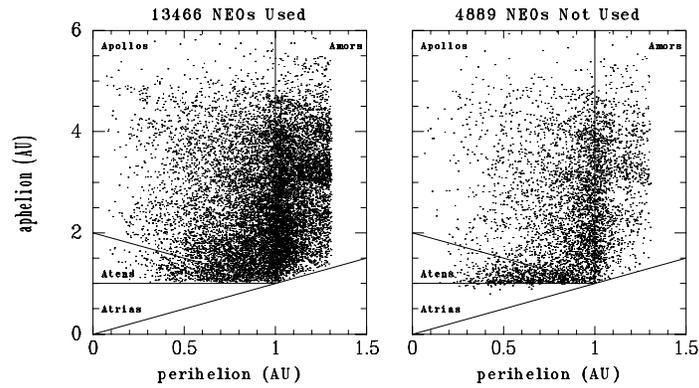}
\caption{
Scatter plot of aphelion vs perihelion distances of the known NEAs.  The
left and right plots separate the NEAs used and not used in the study.  The
criterion for being used are the program sets with large numbers of
observation obtained within 30 degrees of opposition.  The standard
sub-classes of NEAs are indicated.  These illustrate the known distribution
of NEA orbits, that the fraction used in this study is large, and that those
not used are not qualitatively different than those used.
\label{orbits.eps}}
\end{center}
\end{figure}

\begin{figure}[h]
\begin{center}
\includegraphics[width=\textwidth]{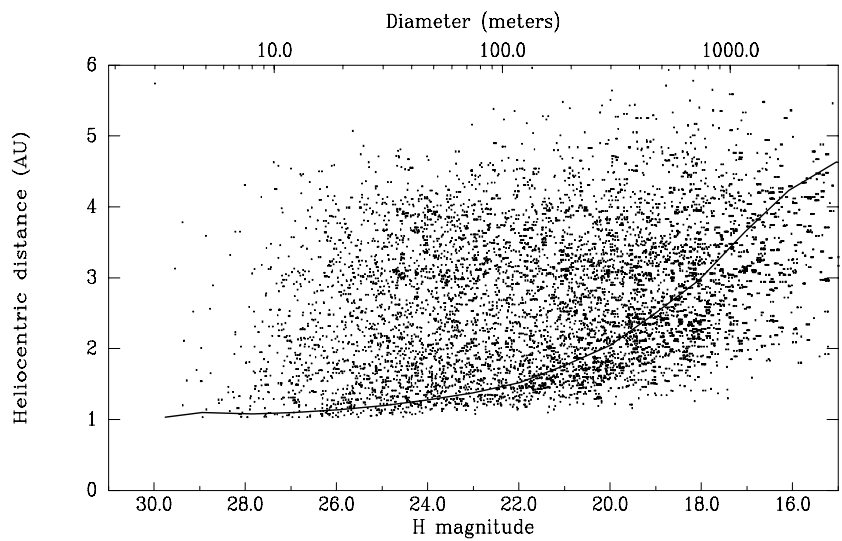}
\caption{
Scatter plot of aphelion distances for NEAs observed by the F51w program
(Pan-STARRS w-band).  In this plot a NEA, represented by each point, will
traverse a vertical line down to it perihelion distance during its orbit.
The solid line is the upper limit boundary of distance at which NEAs were
observed by F51W.  NEAs were only observed at distances below the line
demonstrating that many NEAs will only be observable for a small part of
their orbit in addition to the chance that they appear in the field of
view of the program.  Other programs are qualitatively similar with detection
boundaries varying slightly by magnitude limit.
\label{dist.eps}}
\end{center}
\end{figure}

Figures~\ref{orbits.eps} and~\ref{dist.eps} illustrate this point.  These
show the distribution of orbital and $H$-magnitude parameters for the known
NEAs.  In figure~\ref{orbits.eps} the aphelia distances for the known NEAs
are shown as a function of perihelion distance and in figure~\ref{dist.eps}
as a function of absolute magnitude.  What can be seen in these scatter
plots is that aphelion distances extend to 5 AU and beyond. In
figure~\ref{dist.eps} the solid line is particularly telling in showing the
boundary of distances at which NEAs are observed (in this case by
Pan-STARRS). This illustrates that NEAs are only observable for a small part
of their orbit.

\subsection{Assumptions \label{assumptions}}

Since what is being done is estimating what is not observed from what is known
some assumptions must be made.  The believability of the result depends on
the reasonableness and number of assumptions.  The method presented here
is based on a very small set of assumptions that are hard to dispute and are
almost axiomatic.

The most important assumption is that \emph{the population of known NEAs is
representative of the true population} that we are trying to estimate.  This
includes the range and relative frequency of orbital and $H$-magnitude
parameters in the distribution of known NEAs.  It seems almost axiomatic
since how are we to estimate something that is completely unknown?

The second assumption is that \emph{the orbital phase of the NEAs in the true
population is uniformly distributed}; in orbital parlance, the mean anomalies
are uniformly random.  In plain words, a NEA is equally likely to be found
anywhere along its orbit independent of any other NEA or time of
observation.

The final assumption, that is also almost axiomatic, is that \emph{using the
largest set of observations possible averages out a variety of sins}.  In
particular, the variability of magnitudes, due both to physical rotation and
measurement errors, is averaged out.  In the data used in this study
NEAs typically have multiple observations by the same program and by
multiple programs and at a variety of epochs.

A key factor in this study is that it is specifically about the shape of the
distribution and whether a power law is sufficient to describe the shape.
The absolute numbers are set by a normalization to the biggest NEAs that are
considered to be 95\%, or some similar number, complete.  The importance of
this is that most of the concerns about differences in programs are subsumed
by the normalization.

\subsection{Input Quantities and Method Parameters \label{quantities}}

For all (ground-based) NEA programs the primary observational constraint on
the volume searched is their apparent magnitude limit.  This is almost the
\emph{only} parameter in the method presented in this paper.

The basic input to the method are the measured quantities of magnitude and
time of an observation.  In principle, any observation of a NEA can be used
provided there is an estimate of the magnitude limit for the observation.

The remaining input parameters are the orbital parameters which provide a
distance for the observation and the absolute magnitude $H$ for the NEA.  Note
that these are also observational quantities in that they are derived
directly from reported observations and the axiomatic laws of orbital
mechanics.

The magnitude limit of a particular observation is a fuzzy quantity and is
variable due to weather conditions.  However, the principle that over
a large body of observations there is a typical limit to which the
statistics converge applies.  The effect of uncertainties in this
magnitude limit can and is explored by applying the the analysis with
a range of values.

There is also the question about detection probability for a particular NEA.
It will be shown that this is a small effect.

In this version of the method, observations are limited to be near
opposition.  This is in order to apply a simple inverse-square scaling for
magnitudes. In principle the method can be extended to all observations with
the application of a geometric correction for illumination phase.  However,
it is conceptually and analytically simpler to make this restriction.  It
also the case that most NEO detection and recovery programs primarily
observe near opposition for the obvious reasons of maximum brightness and
use of the full night.

\subsection{Method in Words \label{words}}

Each NEA observation is considered a sample from the class of all NEAs with
nearly identical orbital and absolute magnitude properties.  In this class
all positions along the orbit are assumed to be occupied with a uniform
probability distribution in time.  We also assume that there is no preferred
orientation of the orbit (i.e. the longitude of the ascending node and the
argument of periapsis).  The inclinations are not uniform but what is
required is that the observed NEAs sample the inclinations in the same way
as the full population would be sampled in the absence of magnitude limits.

The assumption that there is no preferred orientation is important.  This
means we can conceptually orient each point along an orbit so that it is
within the same observation window as the specific representative NEA.
In other words, each potential NEA distance from the full population
is from a different orientation of the same ellipical orbit and this is
equivalent to sampling each point in the orbit at different times.  This is
a stationary stochastic approach where orientation is traded with time.
This concept allows evaluating the probability of NEAs with similar orbits to
be observed at different distances and, therefore, different magnitudes,
rate of apparent motion, and time spent within and outside the window of
observability.

Each NEA is generally observed multiple times and usually by multiple
programs.  Given the time of observation and the orbital parameters a
geocentric distance is computed.  Given the apparent and limiting magnitude
of the observation, the maximum observable geocentric distance is computed.
This is purely an inverse square law scaling calculation and the albedo does
not enter.  If the analysis is limited to observations near opposition,
which we do in this paper, any phase effect can be ignored.  This is a key
point that NEA size or albedo don't need to be known and everything related
to apparent magnitudes simply scale geometrically.

As long as the sample also includes all the absolute magnitudes with their
representative orbits, there is also no inherent dependence on $H$ in the
method.  This is key to the primary goal of determining the shape of the HDF.
All the other factors which are not magnitude dependent end up in the
normalization set by asymptoting to the nearly complete bright and large
NEAs.

\subsection{Method in Figures \label{figures}}

Figure~\ref{orbexample} illustrates the key concepts.  The
ensemble of similar orbits (a) shows that when we
describe a NEA as being found anywhere in its orbit and observed at
opposition it is not the exact same orbit but one whose orientation
($i,\Omega,\omega$) is such as to be in the opposition cone.  The
observable and unobservable volumes and the opposition cone are shown
by the broken lines.  In the cartoon one NEA is shown in the
observable volume and three outside but bounded by the aphelion of the
orbit.

The second diagram (b) shows the distribution of heliocentric distances for
the orbit when NEAs are equally likely to be anywhere in their orbit.  As
just mentioned, no single NEA traces a radial distribution of distances
observed at opposition but it is the ensemble of like NEAs that would have
this distribution.  The limitation to near opposition allows interpretation
of the heliocentric distances as geocentric distances with magnitudes (for
an identical NEA) scaling by the inverse square law without worrying about
the llumination phase.  The dashed lines correspond to the Earth's orbit and
the maximum observable distance, for a given observed magnitude, set
by the magnitude limit of the observation.

As in the figure, the time weighted 1D distance ratio is determined by
computing a list of geocentric distances sampled uniformly in time. Adding
up the number of points between the Earth and the distances defined by the
aphelion and limiting magnitude and taking the ratio gives the time weighted
radial factor.  In the example the percentages shown are the fraction of the
time, equivalent to the fraction of NEAs in the ensemble, which are within
the regions.  The ratio of the precentage outside the Earth to what is
observable in this example, $71\%/26\%$, is the 1D volume factor.  One can
think of this as a stretched ratio of the distance in the opposition cone.
As a purely geometric behavior, the cube of this 1D ratio gives the desired
volume ratio.  Note that geometrically it doesn't really matter if the
observing program is truly a cone.

\subsection{Method in Equations \label{equations}}

For each observation $i$ of NEA $j$ by subsample $k$ of program $p$
near opposition, the distance from the Earth, $d_{ijk}$, at which it was
observed is computed from the time of observation and its derived orbital
parameters.  The concept of subsamples is used to evaluate the uncertainties
by considering the scatter in the results across the subsamples in a program
and across all programs.

In this derivation $d$ denotes the distance from the Earth and $r$ is the
heliocentric distance.  By restricting the observations to near opposition
we use the approximation $d=r-1$ where we consistently use AU as the
distance unit.  Note that a \emph{program} is defined to be with a single filter
and consistent observational protocol.

The inverse square law and the observed apparent magnitude $m_{ijk}$ for the
opposition observation allows computing $d(m)$ and $m(d)$ for all NEAs having
a similar orbit and absolute magnitude and observed within the same
opposition window.

\begin{equation}
d(m) = d_{ijk} 10^{0.2*(m-m_{ijk})}, m(d) = m_{ijk} + 5 \log (d / d_{ijk})
\label{d(m)}
\end{equation}

\noindent Note that this purely geometric relationship is not dependent on
any assumptions about albedo or the filter.  Of special significance is
$d(\tilde{m}_{ijk})$ which is the distance to which a similar NEA could be
observed with the limiting magnitude $\tilde{m}_{ijk}$ for the observation.
In the application of the method we make the simplification that all
observations of all NEAs by a particular program are the same and so the
limiting magnitude is $\tilde{m}_k$.

Another magnitude dependent factor is the detection probability
distribution, $P_k(d(m)) = P_k(m(d),v(d))$ where $v(d)$ is the apparent rate
of motion which is a property of the orbit largely governed by the distance.
As with the limiting magnitude this is simplified by assuming $P$ is the
same for all observations and NEAs of the same program.  In the absence of
information about the detection probability distribution we use $P_k \equiv
1$.  The consequence of this is discussed in \S\ref{results}.

\begin{figure}[h]
\begin{center}
\mbox{
  \begin{subfigure}{0.35\textwidth}
    \includegraphics[width=\textwidth]{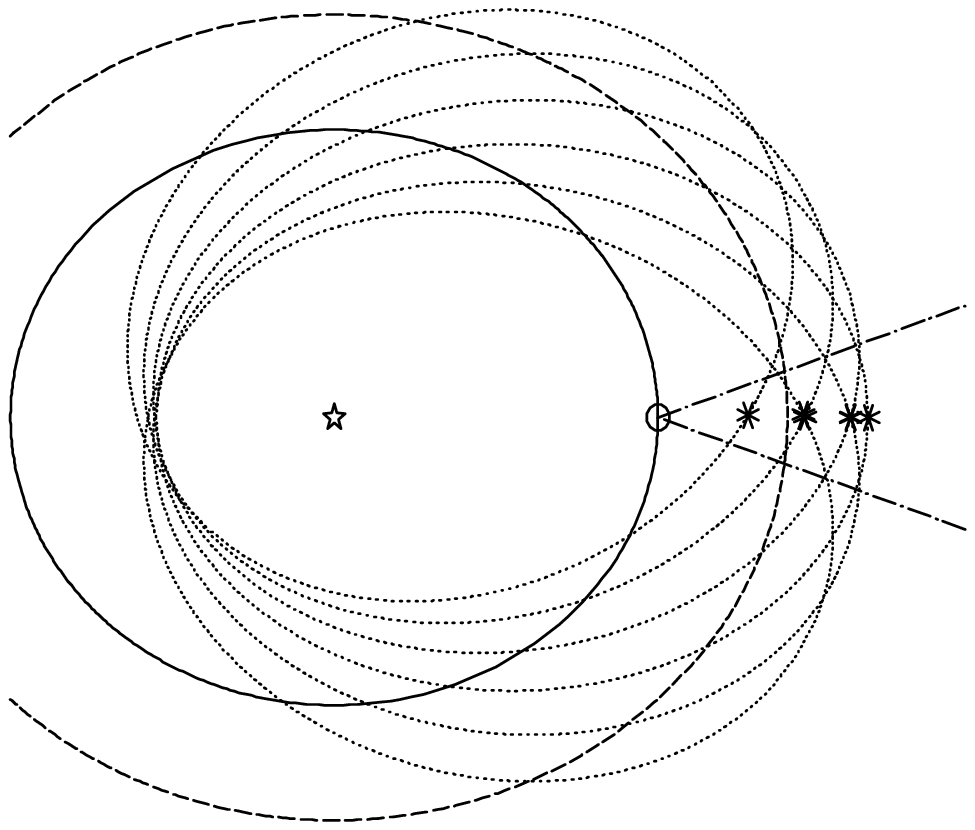}
    \caption{}
    \label{orbalign.eps}
  \end{subfigure}
  
  \begin{subfigure}{0.5\textwidth}
    \includegraphics[width=\textwidth]{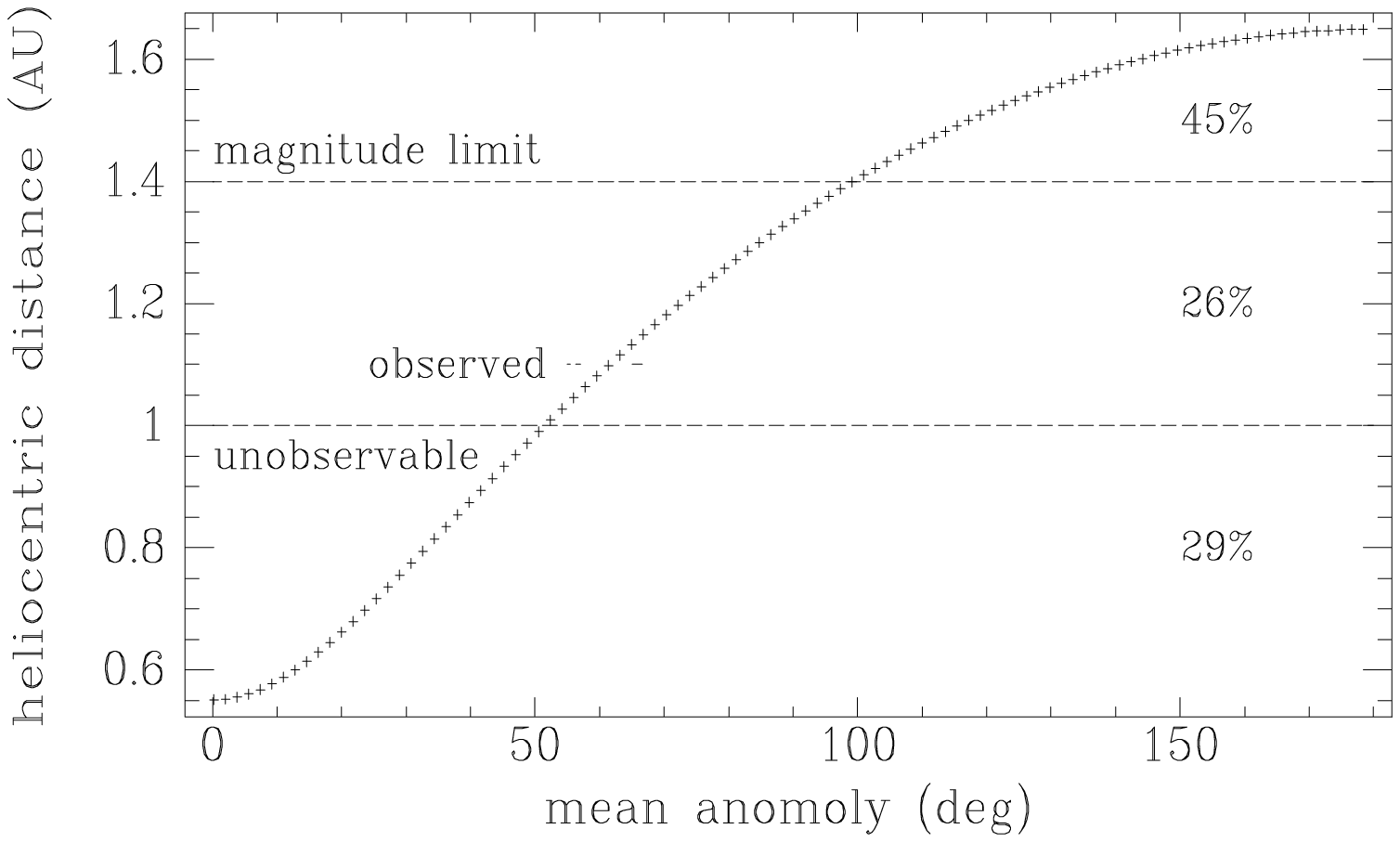}
    \caption{}
    \label{orbexample.eps}
  \end{subfigure}
}
\caption {
These diagrams illustrate the key concepts of the method.  The example
shows a representative NEA with $H$=21.8, a=1.1AU, and e=0.5 observed at a
geocentric distance of 0.1AU and apparent magnitude of 22 by a program with a
limiting magnitude of 22.5.
(a) An ensemble of similar NEAs with the same orbital parameters and
absolute magnitude observed within an opposition cone.  The dashed line is
the magnitude limit.
(b) The heliocentric distances for the ensemble.
The points are evenly sampled in mean anomaly which, by
definition, is equivalent to evenly sampled in time.  The density of points
demonstrates that the NEAs are more likely to be near aphelion and the
percentages are the fraction of time spent in the three regions (interior
to the Earth, observable to the magnitude limit, and unobservable beyond
the limiting magnitude).
The figures are interpreted further in the text.
\label{orbexample}
}
\end{center}
\end{figure}

To estimate the number of NEAs with these orbital parameters consider the
family of orbits where such a NEA is uniformly distributed along the orbit
and, at each point, the orbit is oriented such that the NEA would appear
near opposition.  This is described by the function $t(d(M))$ where $d(M)$
is the geocentric distance of a NEA as a function of the mean anomaly $M$,
and $t(d)$ is the time spent at that distance.  The importance of this
is that uniformly distributed in mean anomaly is equivalent to uniform
in time around the orbit but the time spent at a particular
heliocentric distance is not uniform; asteroids spend more time near
aphelion than perihelion.  As noted previously, near opposition
heliocentric and geocentric distances are simply related.
Figure~\ref{orbexample.eps} provides a particular example of
$t(d(M))$ by considering the density of points in distance where the
points are uniformly spaced in mean anomaly.

For this uniform distribution in mean anomaly, arranged such that
$d=r-1$, the integral of the time spent at each distance over the observable
part of the orbit relative to the whole orbit contributes to the
volume correction factor.

\begin{equation}
z_j = \int_{d_{min}}^{Q-1} t(d_j(M))dM /
\int_{d_{min}}^{d_{max}(\tilde{m})} P(d_j(M)) t(d_j(M))dM
\label{z_j}
\end{equation}

\noindent where $d_{min}$ is the minimum distance for observing the NEA,
$Q-1$ is the maximum distance for the orbit ($Q$ = aphelion distance), and
$d_{max}(\tilde{m})$ ($=\min[d(\tilde{m}),Q-1])$ is the maximum distance a
program could detect a NEA.  While $d_{min}$ is ideally zero it is a
parameter because detections very near the Earth are highly uncertain in
magnitude and extrapolation to the entire orbit is problematic.

Equation~\ref{z_j} is expressed in this somewhat indirect way to motivate
the way it is calculated numerically in application. A table of discrete
distances in uniform steps of $M$ is computed from Kepler's formulas.  This
is conceptually like a large sample of NEAs placed along the orbit in
uniform steps of mean anomaly.  The integrals are then sums of this
population over the distance limits and the ratio is the number potentially
observable verse the number tabulated.  Again, fig.~\ref{orbexample.eps}
illustrates this.

A way to think of $z$ is that it is a non-linear stretch of the volume that
accounts for the variation of the time a NEA spends within the visible and
invisible portion of its orbit rather than simply using using the geometric
distance ratio $z_{geo} = (Q-1-d_{min}) / (d(\tilde{m})-d_{min})$
Equation~\ref{z_j}  reduces to this geometric factor if $t$ and $P$ are
constant. Another way to think of it is as the fraction of time a NEA is
observable with a magnitude limit $\tilde{m}$ relative to an infinitely deep
magnitude limit with a perfect detection probability ($P=1$).

Taken as a stretch in the distance of the observation cone the volume
factor for an instance of a NEA is then
\begin{math}V_{ijk} = (z_{ijk})^3 \label{Vcor}\end{math}.

There is one final factor to consider.  The fraction, $f_j$, of known NEAs
observed by a program for each asteroid type in the population.  For
example, if a program sample included 100\% of the known $H=17.3$ (${\sim}1$ km)
asteroids but only 10\% of the $H=22$ (${\sim}100$ m) ones then $f=1$ and
$f=0.1$ respectively.  This factor allows each program subsample to estimate
the result for the same complete known NEA catalog.  Note this factor is
different than $P$, the probability of detecting an asteroid at different
distances given an observation.

Another way to understand $f$ is that if a program had the sensitivity to
reach the entire orbital volume of every NEA that it observed, where all
$V_{ijk}$ would be one, then the computed distribution would converge to the
known NEA population.

The estimate of the cumulative absolute magnitude distribution, $C$, for a
program subsample is given by

\begin{equation}
C_k(H) = \left\{ \sum_j^{H_j<H} \left( \sum_{i} V_{ijk} \over N_{jk} \right) /
             f_k(H_j) \right\} / A_k
       = \left\{ \sum_j^{H_j<H} \overline{V_{jk}} /
             f_k(H_j) \right\} / A_k .
\label{C_k}
\end{equation}

\noindent The inner sum is the average of the estimates for a particular NEA
from multiple observations and recoveries of the NEA in a program subsample;
which we denote as $\overline{V_{jk}}$.  The observations themselves may
already be an average for each night.  The outer sum is the cummulative
value over the absolute magnitudes where each NEA $j$ has an assigned $H_j$
as part of their catalog parameters.

$A_k$ is a normalization characteristic of the program related to the
effective volume (in space, time, effective depth, etc.) as well as the
detection probability normalization.  It is determined by reproducing the
known distribution of bright/large NEAs in the program where the
incompleteness approaches zero.  The fact that there is this arbitrary
normalization is why the principle, observationally derived, quantity in this
study is the shape of the absolute magnitude distribution.  However, a well
constrained shape tied to the known distribution does provide a well
constrained absolute distribution and most studies of the NEA population
ultimately have some similar normalization to the known bright NEAs.

There are various ways equation~\ref{C_k} could be applied.  In this study
we will compute $C_k(H_m)$ where $H_m$ are in discrete bins.  As will be
shown, over the biggest programs the $C_k(H_m)$ are very similar.  While one
could, in principle, use all the $\overline{V_{jk}}$ in a sum across
programs in eq.~\ref{C_k}, in this study a final absolute magnitude
distribution is derived by averaging $C_k(H_m)$ ($=\overline{C}(H_m)$)
across a program or over all programs at each bin $m$ and using the standard
deviation to evaluate the uncertainties.

\section{Application \label{application}}

The method for estimating the absolute magnitude frequency distribution
described in the preceding section is applied to the catalog of known NEAs
available from the Minor Planet Center in June 2018.  In \S\ref{data}
the statistics of the available data and the selection of programs and
subsamples is presented.  In \S\ref{results} details of how the method
was applied and the results are described.

The believability of this study is gauged by considering the robustness and
uncertainties of the result along with Occam's razor.  The discussion notes
variations that were tried; all leading to the same answer.  Also
\S\ref{simulation} considers a simple experiment demonstrating the result
obtained is not preordained by the method regardless of the input
information.

The analysis was all done in SQL (structured query language) with a database
constructed from all the observations and parameters of the known NEAs.
This allows many tests to be done and repeating the analysis when
the catalog of known NEAs expands over time.

\subsection{Data Selection \label{data}}

The catalog obtained from the MPC in June 2018 contains 18,355 NEAs observed
by 1,670 distinct combinations of MPC program identifiers and filter, which
we call a program ID or simply a program. As described previously, the
analysis is simpler if we consider only NEAs observed near opposition,
chosen as within 30 degrees in ecliptic longitude, which gives 16,313 NEAs.
The analysis was also done with a 20 degree opposition limit with identical
results though, of course, with a smaller sample.  Hence, the result is not
sensitive to the exact value of the opposition window.

\begin{figure}[h]
\begin{center}
\includegraphics[width=\textwidth]{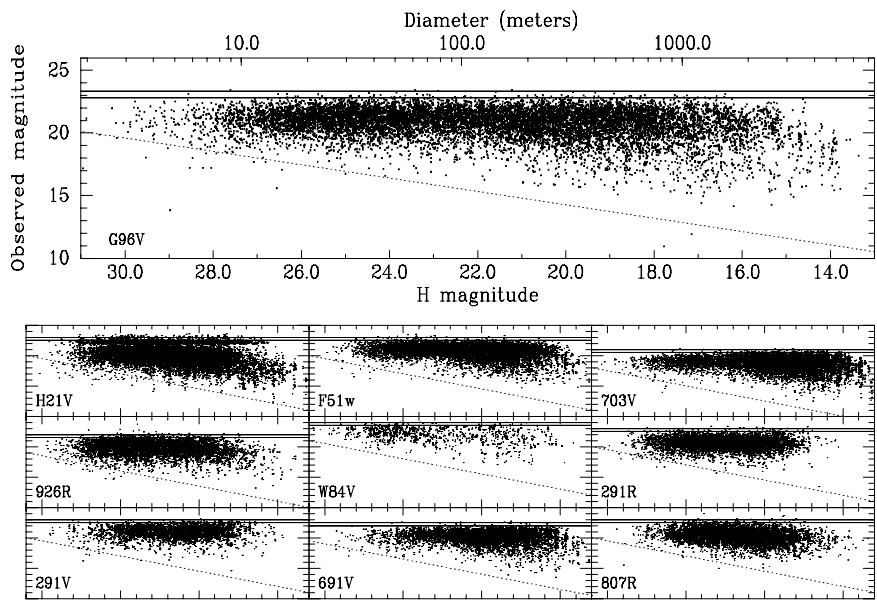}
\caption {Distribution of observed magnitudes as a function of NEA
absolute magnitude for programs included in this study.  The solid lines at
the upper boundary of apparent magnitude are the two magnitude limits
(separated by 0.5 mag) adopted for each program.  The dotted lines are just
indicative of an expected trend that smaller NEAs are observed closer with a
corresponding smaller volume such that they are rarely observed with bright
apparent magnitudes.  The lines have the same slope and are a fixed distance
from the magnitude limit.\label{mags.eps}}
\end{center}
\end{figure}

We further select a limited number of programs for this initial analysis.
There are two reasons for this.  First, a small number of well known
programs account for the majority of NEA discovery/recovery
observations.  Second, while in principle any observation can be used in the
analysis, the requirement that each observation have an associated magnitude
limit leads to programs with a large number of observations allowing solid
estimates of magnitude limits as demonstrated later by figure~\ref{mags.eps}.

The programs were chosen by sorting the number of distinct NEAs observed by
each program in the catalog and selecting 9 of the top 13 programs.  In
addition, the DECam NEO Survey of \citet{NEOSurvey} was included because of
its greater depth and the availability of detection efficiency curves to
check if these are important to the result.  The sorted list is shown in
table~\ref{tab_datasets}.  This also shows the number of nights with
observations N$_\mathrm{night}$.  There are normally 3 or more measurements
of a NEA over a short time span in a night and in this analysis those
measurements are averaged to define a single "observation-night".  The
consideration is that the observed distance and magnitude, the primary
inputs to the method, do not change significantly in these reports and so
give a better estimate for a particular epoch of observation.

Table~\ref{tab_datasets} also shows the 10 programs are divided into two
sets where set 1 is the top 5 programs plus the DECam NEO Survey and set 2
is the remaining selected programs.

The net result of these selections is that 13,466 NEAs from 10 programs
with 74,696 observation nights are included in this study.  This
achieves a key goal of this study to use a substantial subset of all
known NEAs and NEA observations.

A subset of this dataset restricted to Earth crossing asteroids (ECAs) was
also analyzed with identical results, though we don't present specific
figures.  This subset of Apollo and Aten asteroids consisted of 7,334 (out
of 10,839) with 36,541 observation-nights.

The distribution of the selected NEAs in $(q,Q)$ space, perihelion and
aphelion distance, is shown in figure~\ref{orbits.eps} for those selected
and those not selected.  These figures speak to several points made in this
paper: 1) a large fraction of the known NEAs are used and 2) there is no
qualitive difference between the NEAs used and not used.  The second point
relates to the key assumption that the NEAs in the analysis form an unbiased
sample of the known and full population of NEA orbits.

The key, and almost sole, parameter in the analysis is a magnitude limit,
$\tilde{m}_k$, for each observation.  The advantage of programs with large
numbers of consistent observations is that a magnitude limit can be
estimated by looking at the distribution of observed apparent magnitudes
reported.  Figure~\ref{mags.eps} shows the observed and $H$-magnitudes for
each program.  The key features of these plots are that all programs,
regardless of telescope aperture, sample the range of $H$-magnitudes and that
the upper limit of apparent magnitudes has a reasonably clear boundary.
This boundary defines a magnitude limit with little dependence on NEA
absolute magnitude.  In this figure, and all other figures showing scatter
plots of the observations as a function of absolute magnitude, note the lack
of obvious differences in the range $H$=[20:26] where the HSGT studies
find a different character from other parts of the range.

\begin{figure}[h]
\begin{center}
\includegraphics[width=\textwidth]{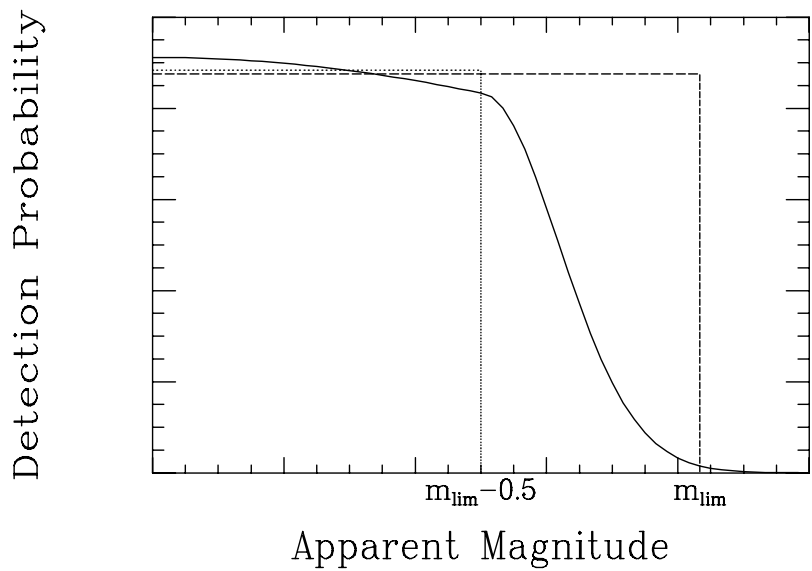}
\caption {Illustration of the relationship between the two magnitude limits
and the simplification of the detection probability $P$ being constant up to
the magnitude limit. If the true probability distribution is bracketed by
the limits then it would be expected that the result using the true
distribution would also be bracketed.  Note that an actual probability
distribution was used for the W84V program in this study.
\label{detprob.eps}}
\end{center}
\end{figure}

The figure shows lines for two magnitude limits used in the analysis.
Table~\ref{tab_datasets} provides the adopted fainter limit (the upper
line).  The second magnitude limit is simply 0.5 magnitude brighter.  Two
limits are used to investigate the robustness of the results with the choice
of magnitude limit and as a way to bound the unknown detection
probabilities (discussed later and illustrated by figure~\ref{detprob.eps}).

Program H21V appears unusual in the figure because two telescopes of
different apertures are used to make observations depending on the expected
apparent magnitude.  We set a magnitude limit, as with the other programs, and
then checked if the analysis showed any pecularities with respect to the other
programs.  The results for just this program were found not to be significantly
different than the other programs.  One could argue that this is expected to be
the case since this is a recovery program rather than a discovery
program.

In order to gauge the uncertainties and robustness of the result the scatter
in many programs and subsamples from the programs are used in the analysis.
In other words, any subsample of the known set of NEAs, providing that each
samples the range of properties ($H$-magnitudes and orbits), should
give the same HFD apart from uncertainties in the measurements and program
techniques.  The scatter in the resultant HFDs provides a measure of the
uncertainties and variations in the data which is a stand-in for the largely
unknowable uncertainties in all aspects of NEA discovery and recovery
programs.

We've already identified two types of subsamples -- by program and by
applying different magnitude limits.  In the results of the following
section we take 20 subsamples of the NEAs observed by a program where each
subsample contains a random selection of half of the NEAs.  Ten of these
subsamples were analyzed with the magnitude limit given in
table~\ref{tab_datasets} and the other ten with a limit that is a half
magnitude brighter.

We note that an exploratory analysis was done with just a single sample of all
NEAs observed by each program and with a single magnitude limit.  The
results reported here with the subsamples are effectively the same but
with a clearer picture of the uncertainties.

What does varying the magnitude limit represent?  As seen in the
detailed derivation of the method there should be a factor for the
detection probability of a NEA at various points along its orbit.
In terms of the observational factors that govern the probabilities
(beyond those imposed by the atmosphere which are stochastic)
these are primarily the apparent magnitude and rate of motion which
directly related to distance.

Detection probabilities generally have a shape the goes from high
probability to zero as the apparent magnitude approaches the magnitude
limit.  Figure~\ref{detprob.eps} illustrates the idea with a cartoon
representation of a detection probability curve in apparent magnitude.
Since we lack detailed knowledge of the detection probabilities for all but
the DECam NEO Survey we are effectively representing the probabilities as a
step function at the adopted magnitude limit.  By using both a brighter and
fainter magnitude limit this is intended to capture the range over which a
program's detection probability function is transitioning.

It was found that use of different magnitude limits and a real detection
probability for W84V (from recovery of embedded synthetic sources) made
little difference to the resulting shape of the HDF in the application of the
method.  For this reason  we don't explore this further.

\subsection{Results \label{results}}

\begin{figure}[h]
\begin{center}
\includegraphics[width=\textwidth]{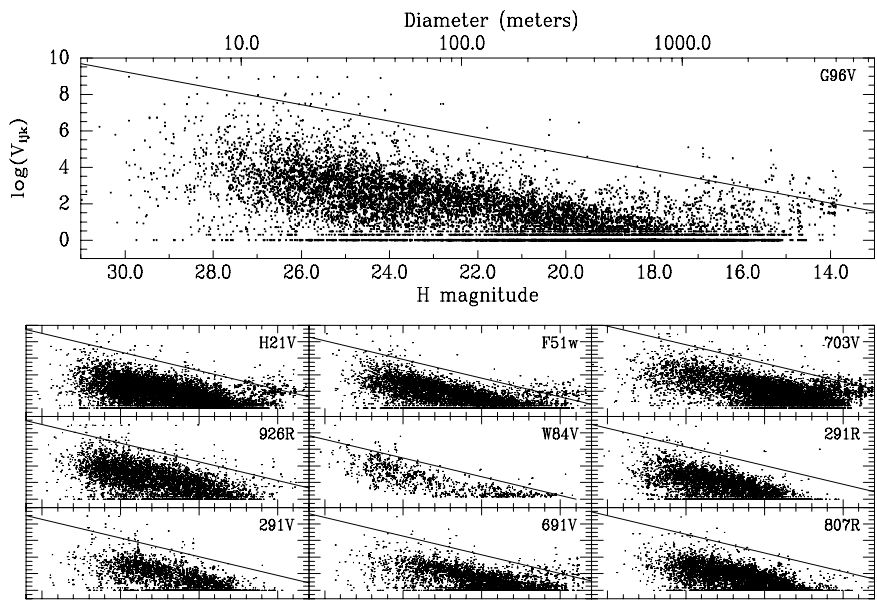}
\caption {Distribution of the volume corrections as a function of
NEA absolute magnitude for programs included in this study.  The solid line
is a limit applied to avoid gross effects from a small number of
observations with either poor absolute magnitudes or extreme ellipticities.
The limits all have the same slope and the origin is tied to the magnitude
limit parameter so that each program is treated the same.  Note that the
curves look qualitatively similar with no obvious differences in the range
$H$=[20:25].\label{vcor.eps}}
\end{center}
\end{figure}

For each observation the analysis consists of computing the volume
correction factor $V_{ijk}$, as defined in section~\ref{equations}, given
the apparent magnitude, magnitude limit, geocentric distance, and orbit
parameters for each NEA observation-night and using a value of 0.02 AU
for the pseudo-parameter $d_min$.  In addition, for program W84V
the known detection probability distribution ($P$ in eqn.~\ref{z_j}) was
applied.  (The detection probability mades only a small difference in the
results showing that it is a minor effect compared to the orbit volume
factor $z$ in eqn.~\ref{z_j}).  Figure~\ref{vcor.eps} shows the factors as a
function of $H$-magnitude separated by program.

The first feature to notice is the scatter plots are fairly similar.  Note
the log scale such that, for example, a value of 4 means that it
represents $10^4$ other NEAs with the same orbit at various distances
including beyond the detection limit near aphelion.

The extreme points seen in fig.~\ref{vcor.eps} produce very large
contributions for a small number of observations.  These points are often
large because of poor or uncertain magnitude measurements (which are
particularly prevalent at the very bright and faint ends) and sometimes due
to large eccentricities with aphelions unusually far away. Such extreme
amplications by a small number of observations have an impact on the HFD
results.  To understand why consider computing an unbinned cumulative
distribution by ordering the points in $H$ and making a point-by-point
cumulative sum.  This generally produces a smooth shape except at the points
with extreme corrections which add large discontinuous jumps.  In between
these jumps the log slope continues on with the same slope as before the
jump.

To minimize this distorting effect of a very few points we introduce an
upper limit to the correction factors as indicated by the lines in
fig.~\ref{vcor.eps}.  We can apply this line either by adjusting the values
down to the limit or by eliminating them from the analysis.  The results
presented here use the former but are essentially identical if the NEA
observations are excluded entirely.  The limit lines are all identical so as to
minimize biases and implicit parameters.  In keeping with having only the
single magnitude limit parameter, the identical lines are simply shifted by
the magnitude limit parameter for each program.

\begin{figure}[h]
\begin{center}
\mbox{
  \begin{subfigure}{0.5\textwidth}
    \begin{center}
    \includegraphics[width=\textwidth]{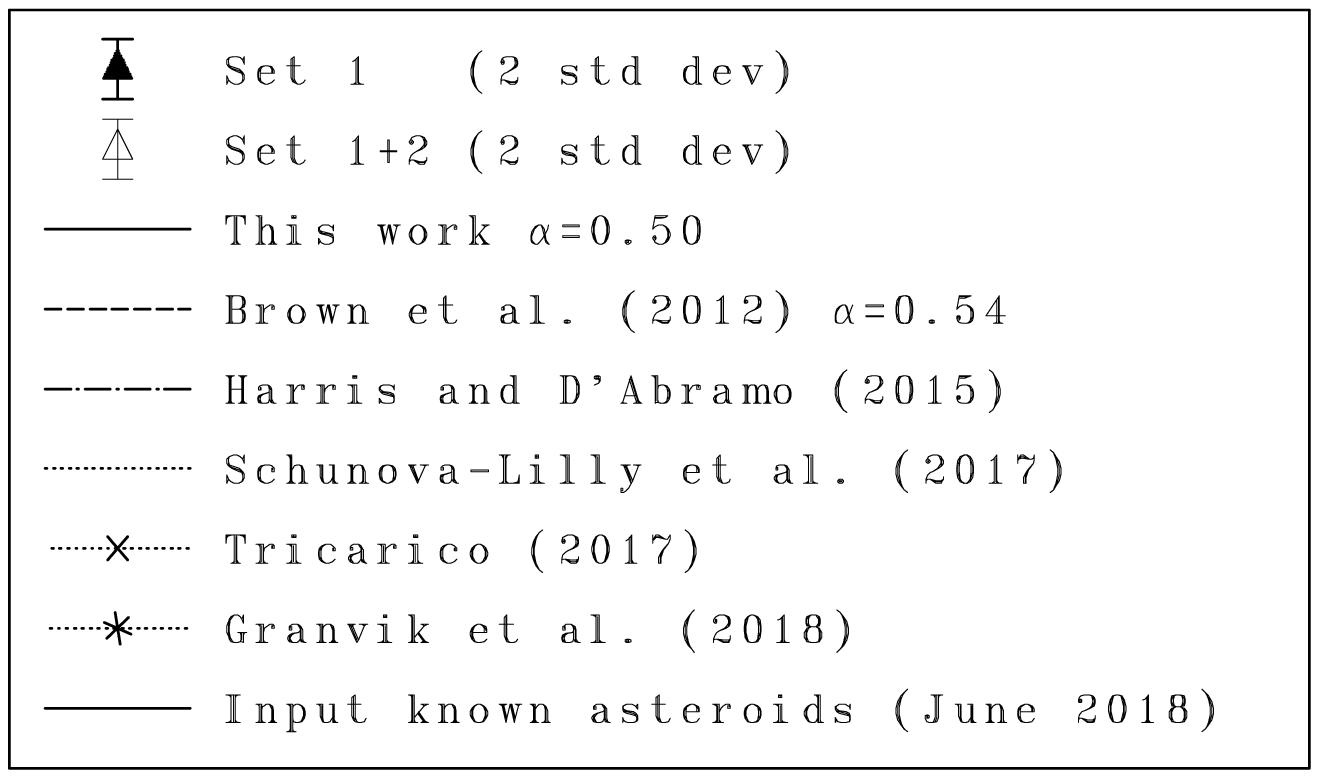}
    \label{main2_legend.eps}
    \end{center}
  \end{subfigure}
  
  \begin{subfigure}{0.5\textwidth}
    \begin{center}
    \includegraphics[width=\textwidth]{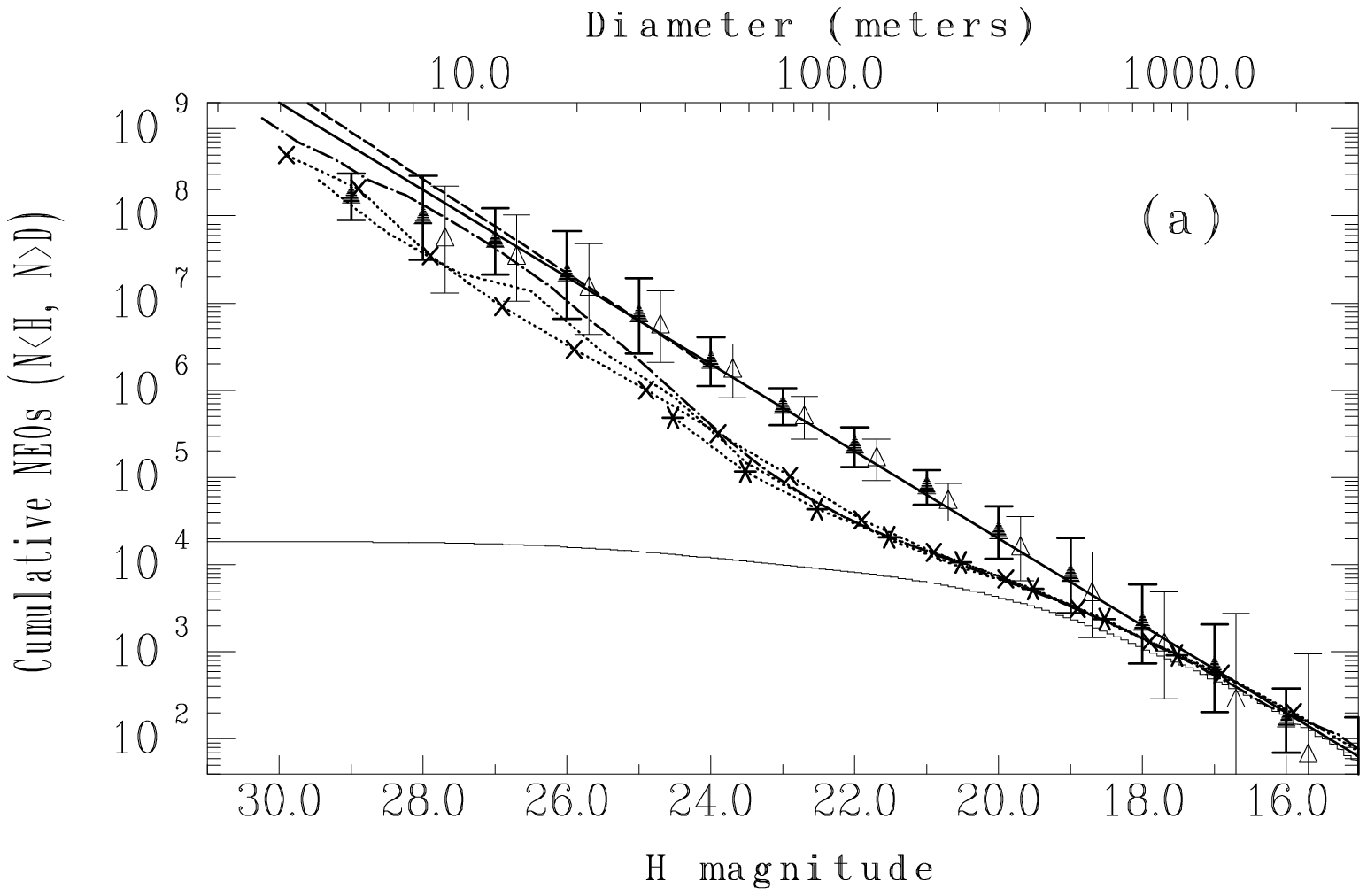}
    \label{main2.eps}
    \end{center}
  \end{subfigure}
}
\mbox{
  \begin{subfigure}{0.5\textwidth}
    \begin{center}
    \includegraphics[width=\textwidth]{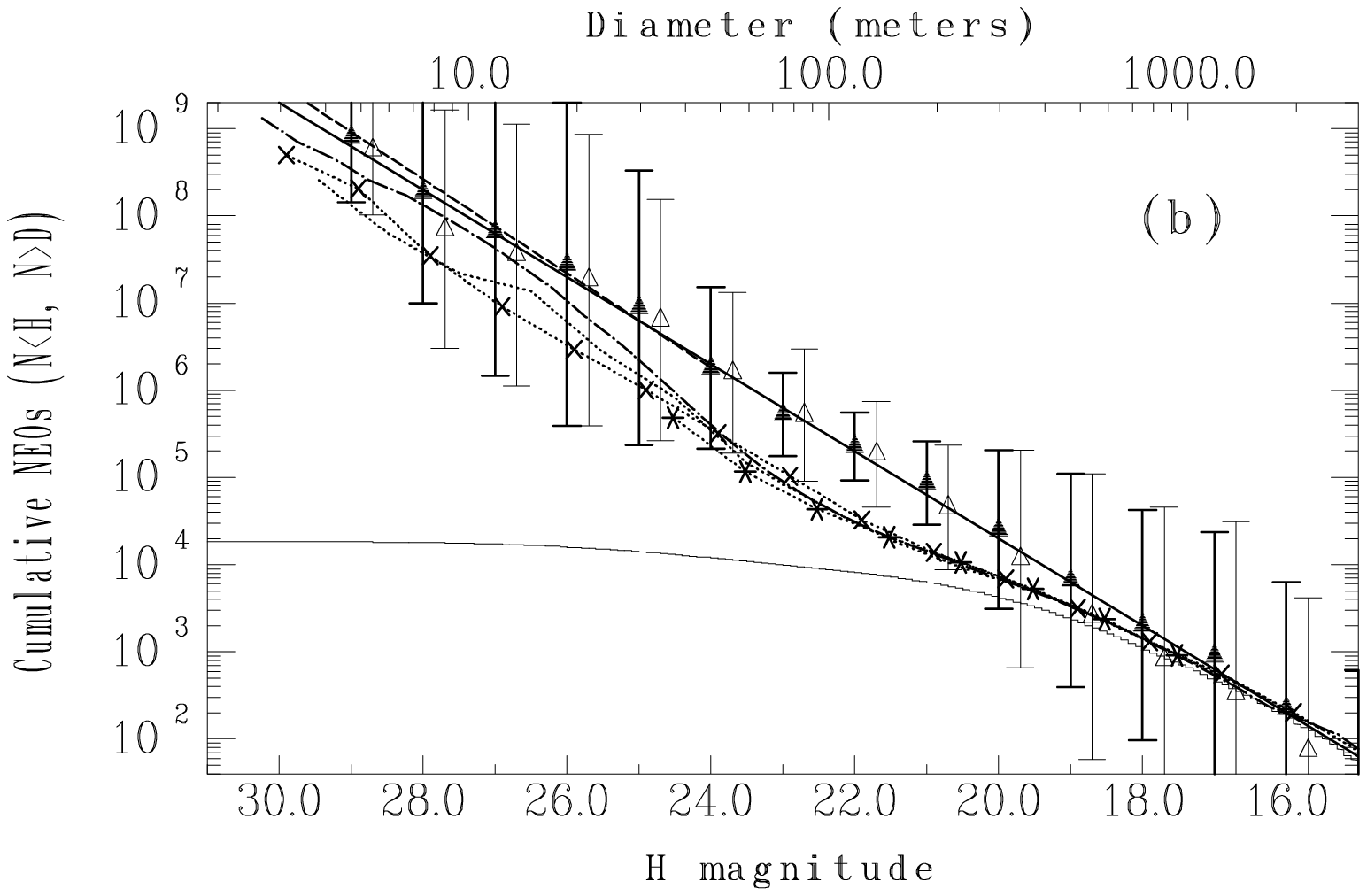}
    \label{main2_novcormax.eps}
    \end{center}
  \end{subfigure}
  
  \begin{subfigure}{0.5\textwidth}
    \begin{center}
    \includegraphics[width=\textwidth]{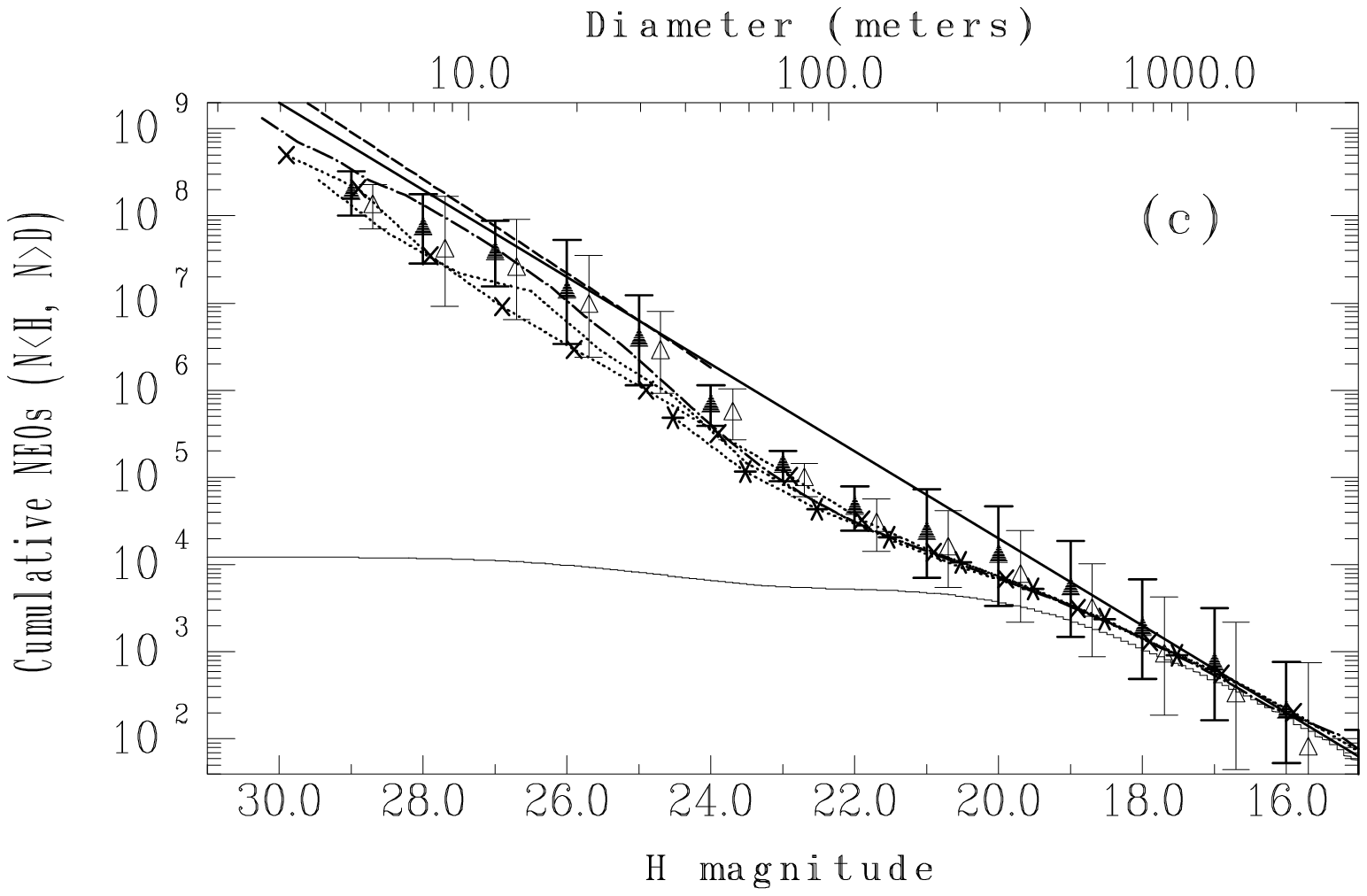}
    \label{main2_sim.eps}
    \end{center}
  \end{subfigure}
}
\caption {NEA HFDs from this and other studies.  The HFDs are normalized to
asymptote to the largest known NEAs.  The heavy solid line is the $\alpha =
0.5$ conclusion of this study. The other HFDs are as indicated in the legend.
(a) Mean and two standard deviation error bars based on multiple subsamples
across multiple programs.  The darker, filled error bars are from the set 1
programs and the lighter open error bars, offset along the $\alpha = 0.5$
line for clarity, are from all programs included in this study. The
contribution from a small number of outlier NEAs have been limited as
discussed in the text and shown in fig.~\ref{vcor.eps}..  (b) The error bars
are computed in the same way as in (a) except there is no adjustment for
outlier NEAs.  (c) The error bars are computed in the same way as (a) except
the input catalog of known NEAs is modified by "undiscovering" some of them
as described in \S\ref{simulation}.  This is a simulation illustrating the
$\alpha = 0.5$ result is not a forgone consequence of the method.
\label{main2}
}
\end{center}
\end{figure}

While justified because these are often magnitude errors, modifying these
small number of points may seem ad hoc.  Therefore, the results are computed
both with and without the limits (discussed later and shown in
fig.~\ref{main2}c).

Each HFD derived from a sample has a similar shape but a different
normalization.  Without making adjustments the scatter between the samples
would be dominated by this.  If one simply normalizes at a particular
$H$-magnitude then the scatter would be zero at that point.  Instead
a power law slope is fit to each sample HFD over the range $H$=[21:25], where
the number of NEAs contributing is largest, and the fit is evaluated
at $H$=22.  All the samples being combined are then normalized to a
log value of 5.3 and the mean and standard deviation is computed
to form the HFD across the combined samples.  Note we are not constraining the
shape of the HFD in this process.

\begin{figure}[h]
\begin{center}
\includegraphics[width=\textwidth]{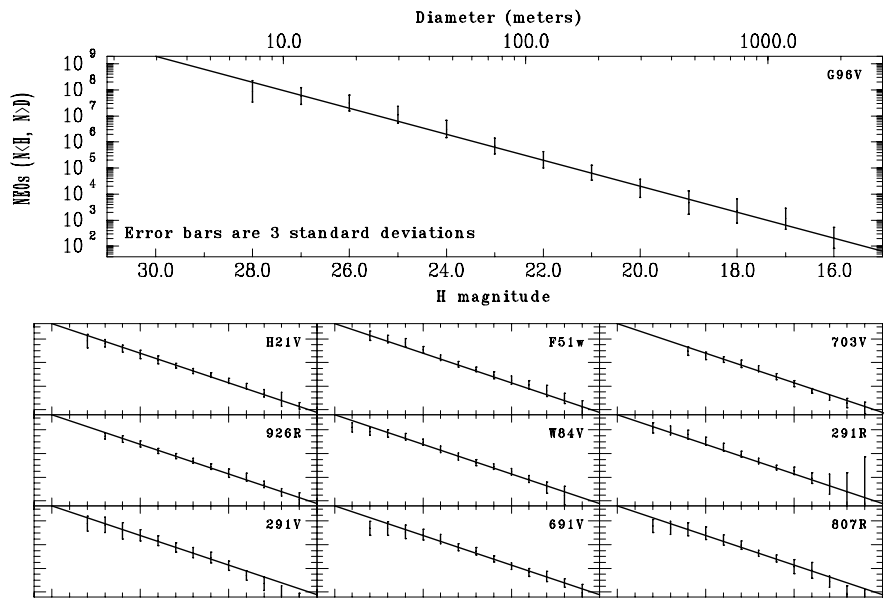}
\caption {The HFD means with three standard deviation error bars for each
program. The solid lines are the standard $\alpha = 0.5$ power law
conclusion of this study. \label{cum.eps}}
\end{center}
\end{figure}

Figure~\ref{cum.eps} shows the HFDs combined by program.  The error bars are
3 standard deviations which essentially describe the scatter in the
samples.  The reference solid line is the same in all HFD plots
presented in this paper and is ultimately the final result we
quote:

\begin{equation}
\log(N_\mathrm{nea}\!\!\!<\!\!\!H) = 0.5 H - 5.7
\label{answer}
\end{equation}

\noindent It is a power law with a slope of one-half and a normalization
that asymptotes to the nearly complete number of large NEAs ($> 2$ km).
The results for each program look very similar and in agreement with
the standard HFD of eqn.~\ref{answer}.

A power law slope is determined for each program using the means and error
bars from the subsamples in the range $H$=[18:27] (a larger range then for the
normalization of the subsamples previously noted).  The values for these
slopes are reported in table~\ref{tab_datasets}.

\begin{figure}[h]
\begin{center}
\includegraphics[height=3.3in]{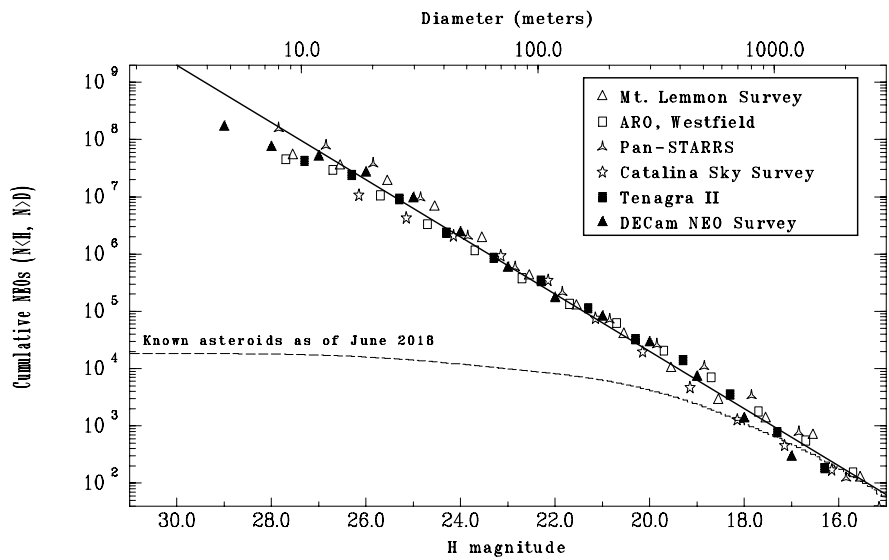}
\caption{The mean values for each program in set 1 are plotted with the
solid line showing the standard $\alpha = 0.5$ HFD conclusion of this study.  The
dashed line is the cumulative distribution of the known NEAs.  For clarity
the positions of the points are shifted along the 0.5 power law, program set
2 is not included, and the error bars are omitted.
\label{main1.eps}}
\end{center}
\end{figure}

The agreement between different programs can be seen in figure~\ref{main1.eps}.
This plots the mean values for the HFDs in program set 1 at each magnitude.
The $H$-magnitude bin positions are shifted along the standard 0.5 power law of
eqn.~\ref{answer} so the values can be more clearly seen.  For clarity only
set 1 and no error bars are shown.  This reinforces the conclusion that
different programs agree well with a simple HFD for the NEAs and we can
proceed to combine all the samples across all the programs.

Figure~\ref{main2}a combines all subsamples from program set 1 (the
darker filled symbols) and for sets 1 and 2 together (the lighter open
symbols).  The points for the two combinations are offset along the standard
line so the two don't overlap.  The error bars are 2 standard deviations.
The power law slopes fit to these combinations are reported at the bottom of
table~\ref{tab_datasets}. The additional lines in the figures will be discussed
later.

The only ad hoc element in this analysis is the upper limit, shown in
fig.~\ref{vcor.eps} and discussed earlier, placed on the volume corrections.
These cause large distortions in the cumulative counts by a very small
number of observations.  To evaluate whether this affects the results, the
analysis was also performed without this limit; i.e no rejection or
alterations to the full dataset including extreme outliers.  The effect of
not applying a limit is an increase in the scatter of the subsamples as
shown in fig.~\ref{main2}b.  This does not constrain the slope of the
distributions very well but it clearly shows a simple power law, consistent
with eqn.~\ref{answer}, is most likely and is not consistent with their
being structure in the HDF of the form seen in HSGT.

\subsection{Undiscovery Simulation} \label{simulation}

A study with population models and simulated observation strategies would be
valuable to see if the method recovers the input population but that is
beyond the scope of this paper.  However, there is a simple simulation that
can address a particularly important question:

\begin{quote}
\emph{Is there is something in the methodology that forces a simple power law?}
\end{quote}

The simulation, which is more in the nature of an experiment, consists of
randomly "undiscovering" a subset of NEAs.  The reason for using an
undiscovery simulation is that no orbits or detection strategies have to be
created.  We are just saying those orbits and magnitudes/sizes don't exist
in the simulated full population and so are not discovered and added to the
catalog of known NEAs.  Any pattern of undiscoveries that is not uniform in
absolute magnitude but blind to orbital parameters can be used to answer the
question.

The pattern we use is a gaussian probability distribution that
preferentially undiscovers NEAs around $H$=22.  The undiscovery probability
distribution used in this simulation is shown by the dotted line and
right-hand axis in figure~\ref{ratio.eps}.  This distribution has a
particular significance as will be discussed later.  Randomly undiscovering
these NEAs produces the simulated known NEA population given by the dashed
line in figure~\ref{mpchist.eps}.  The random undiscovery was over the full
catalog of known NEAs with no selection by orbit, whether a NEA was
observed by a particular program, or within the opposition window.  The
analysis performed on the complete known NEA population (the solid line in
fig.~\ref{mpchist.eps}) is repeated with the simulated known population.
The HFD result is shown in figure~\ref{main2}c.

As seen in the figure, the result is no longer a structureless power law.
Rather it has a dip that is, by design, a good approximation to the HFDs of
HSGT.  Note, however, that the intent of this simulation is to answer a
question about the methodology and not to try and explain the HSGT
structure.  There is no physical basis or claim in the undiscovery
simulation related to why that structure is found by HSGT.

The answer to the question posed in this section is then:

\begin{quote}
\emph{The methodology described in this paper does \textbf{not} force the
simple power law found using the complete catalog of known NEAs!}
\end{quote}

\section{Discussion \label{discussion}}

This analysis, based solely on observations and the catalog of known NEA
orbits derived from them, along with some basic assumptions, leads to the
simple and robust result that the cumulative absolute magnitude frequency
distribution of near-Earth $A^3$ asteroids (HFD) is a power law with the
intriguing slope of 0.5.  The only sign of a departure from this is for the
faintest, smallest, and hardest to find NEAs where the number of known
examples is limited.

It is tempting to just present this result as it stands.  However, the
difference with respect to HSGT immediately draws attention from anyone
familiar with those studies.  This difference is also important for
estimating the challenges in the planetary defense mandate for finding most
of the NEAs larger than 140m.  So we address this "elephant in the room".

Figure~\ref{main2} includes the HFDs from the commonly cited studies.
\citeauthor{Brown} (the dashed) line provides a consistency check with
bolide observations.  That work suggests a power law slope of 0.54 for the
smallest NEAs (those which impact the Earth).  This is not too dissimilar to
the slope found in this study.  The authors indicate that this slope
connects to the slightly larger NEAs.  In the figure the slope is connected
around $H$=25.

Where a significant discrepancy is seen is in the studies of HSGT whose HFDs
are also shown in the panels of fig.~\ref{main2}.  At a coarse level the
overall slopes and tendency to reach similar cumulative numbers for the
smallest NEAs is in reasonable agreement.  However, the notable feature is a
dip in the range $H$=[20:26].

\begin{figure}[h]
\begin{center}
\mbox{
  \begin{subfigure}{0.5\textwidth}
    \includegraphics[width=\textwidth]{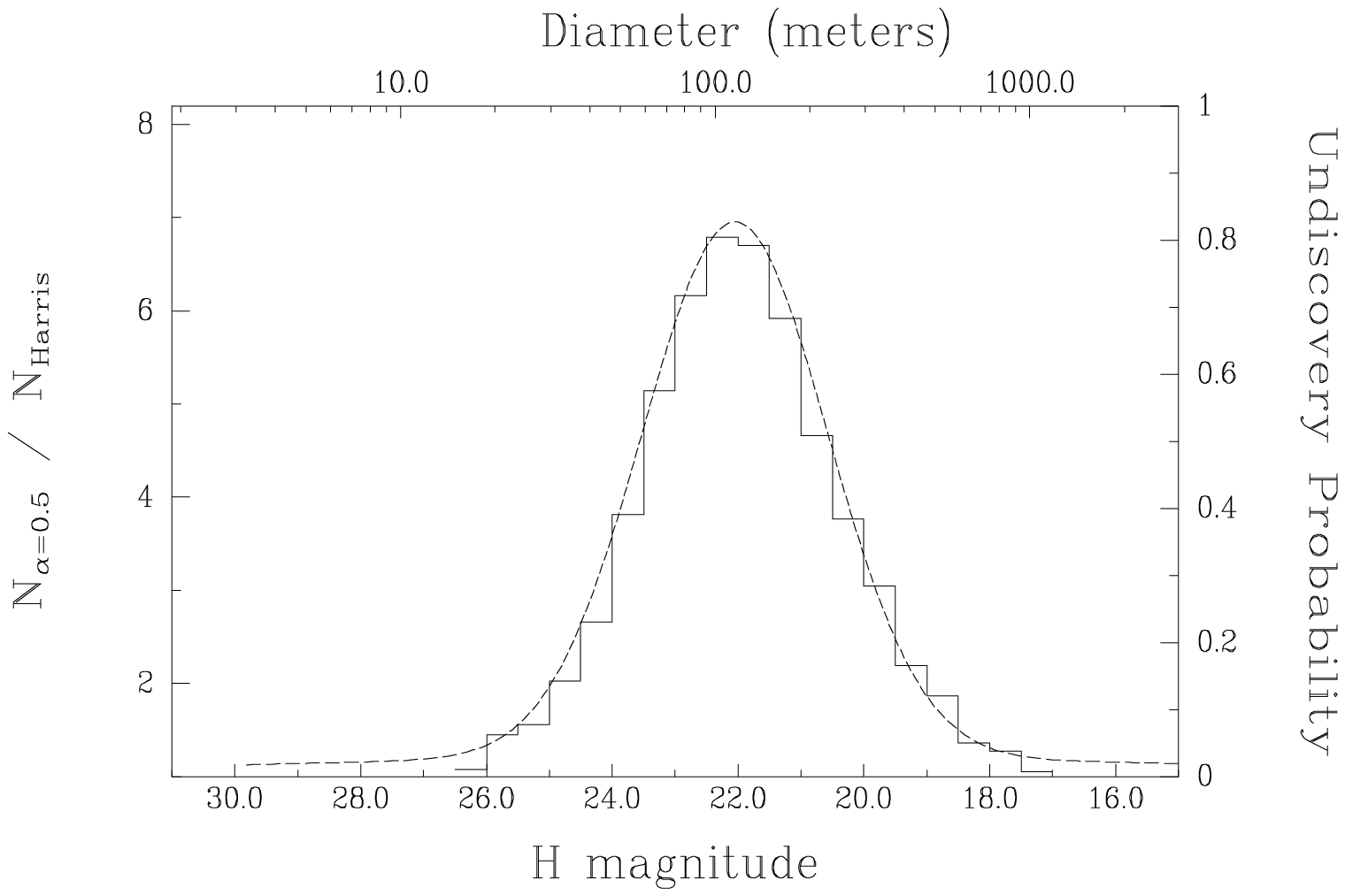}
    \caption{}
    \label{ratio.eps}
  \end{subfigure}
  
  \begin{subfigure}{0.5\textwidth}
    \includegraphics[width=\textwidth]{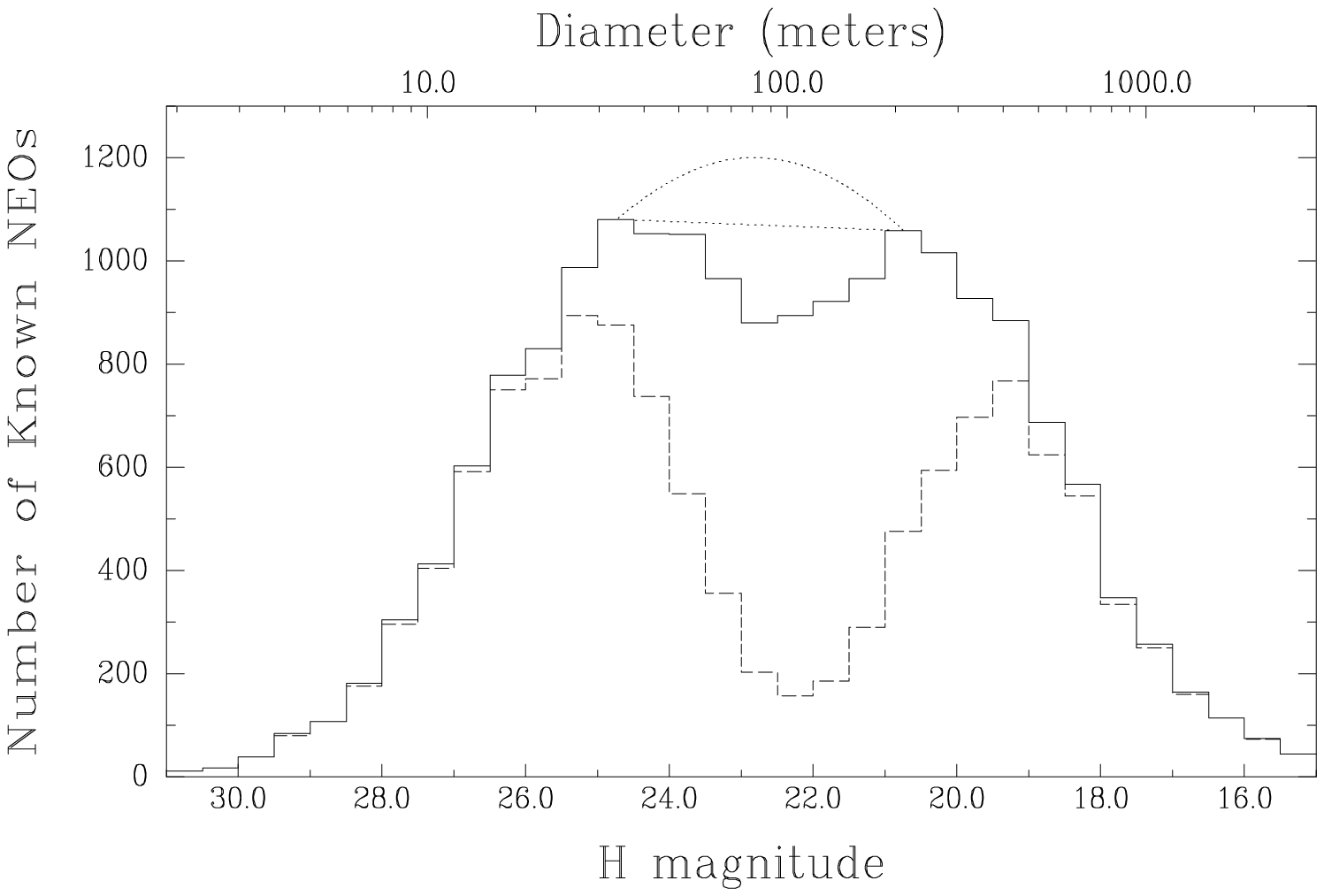}
    \caption{}
    \label{mpchist.eps}
  \end{subfigure}
}
\caption{
(a)~The solid binned line and left axis is the ratio of the
$\alpha=0.5$ (this work) to the \citeauthor{Harris} (non-cumulative)
histograms normalized to be equal at $H$=16.  The dashed line and right axis
is a probability distribution for discovered NEAs to be "undiscovered" in a
simulation.  (b)~The solid line is the histogram of discovered NEAs as of
June 2018.  If one assumes the shape should not be bimodal then the dotted
lines extrapolate a plateau and gaussian for characterizing a possible
discovery deficit.  The dashed line is a histogram with known NEAs
"undiscovered" using the probability distribution in (a) for the simulation
described in \S\ref{simulation}. \label{dip}
}

\end{center}
\end{figure}

To more clearly quantify the magnitude of the difference,
figure~\ref{ratio.eps} displays the ratio (in non-cumulative linear units)
of the \citeauthor{Harris} distribution to the simple $\alpha=0.5$ power law
of this study.  This shows a difference of up to 700\% at $H$=22 ({$\sim$}100m).
This ratio can vary by a small amount depending on the details of the
normalization which is at $H$=16 in this case.  As seen in fig.~\ref{main2},
the \citeauthor{Schunova}, \citeauthor{Granvik}, and \citeauthor{Tricarico}
ratios would be similar.

An argument typically used to justify the reality of this dip is that there
is also a dip in the raw numbers of discovered NEAs in approximately the
same range of magnitudes.  \citeauthor{Harris} claim this makes their result
definitive.  To consider this we can look at figure~\ref{mpchist.eps} that
plots the histogram of the known NEAs.  There is, indeed, an apparent dip in
what might be expected in the range $H$=[21.5:25] if we \emph{assume} the
histogram should look like a single peak or plateau, but it could equally well
be a superposition to two peaks. Many people have speculated on possible
reasons for this histogram as both indicative of something in the true
population (e.g. a rubble pile to monolithic transition) or a discovery
selection effect (e.g. a transition between fast versus slow discovery
methods).

The deficit relative to a plateau or gaussian (shown as extrapolations with
dotted lines) at $H$=22.75 is in the range [22\%:37\%].  While one could
attempt to link this to the factor of 700\% in fig.~\ref{ratio.eps} this
seems to be a somewhat extreme amplification.  While the undiscovery
simulation (\S\ref{simulation}) is not a physical explanation, the
particular undiscovery probability distribution used (fig.~\ref{ratio.eps})
can be considered in the context of this argument.  In order for the method
developed in this paper to find a dip of the size in the HSGT HFDs,
assuming it is caused by a deficit in discovered asteroids,
requires the much more significant deficit shown in fig.~\ref{mpchist.eps}.

So why doesn't the structure around $H$=22 in the histogram of known NEAs affect
the HFD in this study?  The obvious answer is that the method yields
the true population distribution from the sample just as it accounts
for a steadily increasing number of NEAs at small $H$-magnitudes despite
the fact that the discovery histogram is declining there.

One might be tempted to discount this study and give greater weight to the
HFDs of the HSGT studies, as well as the derivative version of
\citet{Stokes}, because the latter studies are similar and outnumber this
work.  This is not the place to review those other studies.  However, there
are notable commonalities in the methodologies, beyond ultimately tying back
to the catalog of known NEAs which all studies must do, which makes them
less distinct than might be apparent.  They all involve generation of a
synthetic population at some point; either by sampling from a subset of
known orbits or making use of the same population models of \citet{Bottke}
and \citet{Greenstreet}.  They also generally involve modeling of detection
efficiencies of the surveys used in their analysis through various means
including simulated surveys over the population models.

The method and result presented here likely raises several additional
questions for the readers.  These commonly asked questions are considered
next.  The answers lie in the three primary assumptions made earlier.

What about the different methods and sensitivities of the various programs
and differences between survey vs. recovery focuses?  As long as the entire
community has discovered a fair sample of the true population the only thing
that enters in the analysis is the observed magnitudes at specific times and
orbital distances and an estimate of the magnitude limit for each
measurement.  Hence any observation could be used.  Large programs, both
survey and recovery oriented, are used simply because it is possible to
estimate a magnitude limit from statistics of the reported measurements.  A
second aspect to this question is that this study is about the shape of the
distribution; i.e. the relative numbers at different sizes.  Most of the
questions about survey methodology, area covered, cadences, etc.  enter into
the normalization which ultimately must tie to the bright end of the
distribution.

What about albedos? As with the previous answer, unless albedo has affected
the representative sample of NEAs in the catalog in some systematic way,
albedo does not enter into the analysis at all.  Whatever the albedo,
and hence apparent magnitude at some distance, the method just makes use
of the inverse square law to find the volume observable up to the
program's limiting magnitude.

What about magnitude variability and rotation?  It is true the magnitudes in
the catalog contain variability, not only from physical effects but from
measurement errors of various kinds.  However, since each NEA is typically
observed several times by different programs at different times, the
principle of large number statistics average out the light curves.

\section{Conclusion \label{conclusion}}

This paper presents a method for estimating the population of near-Earth
$A^3$ asteroids and their absolute magnitude distribution.  It is
solely based on observations and simple assumptions rather than orbital
population modeling.  There are two clear conclusions.

The first is that the cumulative absolute magnitude distribution is a
power law with a slope of $0.50\pm0.03$.  The simple form and
value of the slope is strongly favored by an Occam's razor argument.
There is nothing in the methodology that forces this result.  It is
also extremely robust.

The analysis was also performed with just the Apollo and Aten Earth crossing
asteroids.  The result was an identical power law distribution.  This
shows there is no significant difference caused by Earth crossings.

The second conclusion is that this method does not show the "dip" in the HFD
found by other studies (e.g. HSGT).  The importance of the difference
between this study and those showing the dip is that the size range where
the discrepancy is greatest (around 100m) is what the planetary defense
initiative is hoping to catalog.  A factor of ${\sim}$7 would have a major
effect on the scale of effort and time needed to achieve the discovery
mandate.

\section{Acknowledgements \label{acknowledgements}}

This work is a synthesis of the many years of effort by many teams surveying
the near-Earth asteroid population and the many observatories supporting
them.  It also depends on the compilation work of the Minor Planet Center.

\newpage

\nocite{Brown,Greenstreet,Harris,Schunova,Granvik,Stokes,Tricarico,Bottke}
\bibliography{NEASFD}
\bibliographystyle{abbrvnat}



\newpage

\begin{table}[h]
\caption{Table of datasets. \label{tab_datasets}}
\begin{center}
\begin{tabular}{|l|r|r|c|r|r|} \hline
\multicolumn{1}{|c|}{\textbf{ID}} &
\multicolumn{1}{c|}{\textbf{N$_\mathbf{neo}$}} &
\multicolumn{1}{c|}{\textbf{N$_\mathbf{night}$}} &
\multicolumn{1}{c|}{\textbf{Set}} &
\multicolumn{1}{c|}{$\mathbf{M_{lim}}$} &
\multicolumn{1}{c|}{\textbf{$\mathbf{\alpha}$ [18:27]}} \\
\hline \hline
G96V - Mt. Lemmon Survey         &6890 &12440 &1 &22.5 &0.54$\pm$0.016 \\
H21V - ARO , Westfield           &6252 &14895 &1 &23.0 &0.45$\pm$0.012 \\
F51w - Pan-STARRS                &5379 &8990  &1 &23.0 &0.50$\pm$0.015 \\
703V - Catalina Sky Survey       &4248 &9256  &1 &21.0 &0.51$\pm$0.011 \\
926R - Tenagra II Obs.           &3442 &5813  &1 &22.0 &0.47$\pm$0.012 \\
291R - LPL/Spacewatch II         &3295 &5618  &2 &23.0 &0.53$\pm$0.017 \\
807R - CTIO                      &3091 &6759  &2 &23.0 &0.55$\pm$0.022 \\
J95R - Great Shefford            &2906 &      &  &     & \\
691V - SO KPNO/Spacewatch	       &2699 &5827  &2 &22.5 &0.52$\pm$0.018 \\
204R - Schiaparelli Obs.         &2620 &      &  &     & \\
474R - Mount John Obs.           &2020 &      &  &     & \\
568R - Mauna Kea                 &1832 &      &  &     & \\
291V - LPL/Spacewatch II         &1784 &3728  &2 &23.0 &0.56$\pm$0.022 \\
&\multicolumn{1}{c|}{...}            &      &  &     & \\
W84V - DECam NEO Survey          &512  &1225  &1 &24.0 &0.51$\pm$0.013 \\
&\multicolumn{1}{c|}{...}            &      &  &     & \\
\hline
Set 1                          &     &      &  &     &0.50$\pm$0.026 \\
Set 1 + 2                      &     &      &  &     &0.51$\pm$0.029 \\
\hline
\end{tabular}
\end{center}
\end{table}

\end{document}